\documentclass[twocolumn,aps,showpacs]{revtex4}

\usepackage{graphicx}
\usepackage{dcolumn}
\usepackage{bm}

\begin{document}

\title{Effect of Dimple Potential on Ultraslow Light in a Bose-Einstein Condensate}
\author{Devrim Tarhan$^1$, and Haydar Uncu$^2$}
 \affiliation{$^1$Department of Physics, Harran University, 63300
Osmanbey Yerle\c{s}kesi, \c{S}anl\i{}urfa, Turkey}
\affiliation{$^2$Department of Physics, Adnan Menderes University,
Ayd\i{}n, Turkey}
\date{\today}
\begin{abstract}
We investigate the propagation of ultraslow optical pulse in
atomic Bose-Einstein condensate in a harmonic trap decorated with
a dimple potential. The role of dimple potential  on the group
velocity and time delay is studied. Since we consider the
interatomic scattering interactions nonlinear Schrödinger equation
or Gross-Pitaevskii equation is used in order to get the density
profile of the atomic system. We find large group delays of order
$1$ msec in an atomic Bose-Einstein condensate in a harmonic trap
with a deep dimple potential.
\end{abstract}

\pacs{42.50.Gy, 03.75.Nt} \maketitle
%

\section{Introduction}
The impressive demonstration of ultraslow light propagation
through an atomic Bose-Einstein condensate (BEC)
\cite{slowlight-exp1}, utilizing electromagnetically induced
transparency (EIT) \cite{eit1}, has promised an appealing
application as a quantum memory \cite{liu}. Beside the ultracold
atoms, slow optical pulses have been observed in various media
like hot rubidium vapor\cite{slowlight-hotgas1}. In BECs,
ultraslow optical pulse is delayed by the order of few
microseconds\cite{slowlight-exp1}. After, large controllable time
delays for such broadband pulses were
proposed\cite{shortpulse-prop}. Nevertheless large group delay of
light was observed in hot rubidium gases\cite{slowlight-hotgas1}.
Quite recently, controlling the group velocity has been
discussed\cite{boyd}. It has been theoretically and experimentally
shown that this delay time can be increased by increasing the
atomic density\cite{storedlight}. Moreover, dimple potential can
be used for increasing the time delay since the density of
Bose-Einstein condensate can be increased by means of the dimple
potentilas\cite{pinkse}. Phase space density can be enhanced by an
arbitrary factor by using a small dimple  at the equilibrium point
of the harmonic trapping potential \cite{kurn}. Recently, such
potentials are also proposed for efficient loading and fast
evaporative cooling to produce large BECs \cite{comparat}.

If the atom-atom interactions can not be neglected, the structure
of ground state of BEC is described by Gross-Pitaevski equation
\cite{pita-gross}. If the scattering length $a_s$ is much less
than the mean interparticle spacing, Gross-Pitaevski equation
describes the zero-temperature properties of the non-uniform Bose
gas.

In this paper, we investigate the one dimensional propagation of
ultraslow optical pulse in an atomic Bose-Einstein condensate in a
harmonic trap decorated with a dimple potential which is located
at the center of harmonic potential. We study the role of the
dimple potential on the group velocity and the time delay. We
model the dimple potential by a Gaussian function which has a
narrow width value. Our calculations show that it is possible to
observe large group delays up to 1 msec for an optical pulse in an
atomic Bose-Einstein condensate which is trapped in harmonic
potential decorated with a deep dimple. The paper is organized as
follows: First of all the calculation of density profile of an
atomic BEC system is briefly reviewed. After that, propagation of
ultraslow light under EIT system is discussed. We present our
calculations and  discuss the results in result section. Finally,
we conclude in the last section.

\section{Density profile of a trapped Bose gas with a dimple potential}
The density profile of a Bose-Einstein condensate in an external
potential can be approximated very well by means of Thomas-Fermi
approximation. The density of ground state is given by the
absolute square of the ground state wave funvtion:
$n(r)=|\psi_0(r)|^2$. Thomas-Fermi approximation (TFA) is often a
good approximation to assume the total ultracold atomic density is
in fact constant during the weak light field propagation. On the
issue of the dimple, the validity of TFA depends on the length
scale of the dimple. If it is much larger than the healing length,
then TFA with the trap+dimple potential would work fine. The
healing length or coherence length of the BEC is given by $ \xi
=(1 / 8 \pi n a_s)^{1/2}$ \cite{pethick}. Here $n$ is the density
of the ultracold atomic medium and we can take it as the peak
density ($n=\rho(0)$) of the atomic system. We consider range of
parameters in this work within the range of validity of TFA. One
can approximate total density profile of an ultracold atomic
system for 1D (ground state density) by \cite{pethick} $\rho(z) =
[(\mu-V(z))/U_{0}]$. Here $U_0=4\pi\hbar^2 a_{s}/m$ where $m$ is
atomic mass and $a_s$ is the atomic s-wave scattering length.
$\mu$ is the chemical potential and can be evaluated by using
Thomas-Fermi approximation. The chemical potential is determined
from $N=\int\,\mathrm{d}z\rho(z)$. We consider the dimple
potential which is modelled by a Gaussian function. We represent
one dimensional harmonic potential with a dimple as
\begin {equation}
\label{potential}
 V(z)=\frac{1}{2} m \omega_z^2 z^2-V_0
\exp \{-(\frac{z}{\sqrt{2}l_z})^2 \},
\end {equation}
where $\omega_z$ is the trap frequency of the harmonic trap in the
$z$ direction, and $V_0>0$ shows the strength (depth) of the
dimple trap ($V_d=-V_0 e^{-(\frac{z}{\sqrt{2}l_z})^2}$)located at
$z=0$. It is possible  to increase $V_0$ from $0$ to $1500\hbar
\omega_z$ \cite{proukakis}. We use large values for the width of
Gaussian function $l_z$ (in $G(z)= V_0 \exp{(-z^2/2l_z^2)}$)
compared to the extension of BEC (see Sec IV.) in order to get an
appropriate modeling for the narrow dimple. If we apply a deep
dimple to the atomic condensate in a harmonic trap we will get the
total number of atoms analytically by:

\begin {eqnarray}
 N&=&\int_0^{\sqrt{\frac{2(\mu+V_{0})}{(m \omega_z^2+V_0/l_z^2)}}}\,\mathrm{d}z \frac{1}{U_0} \nonumber \\
 &\times& \{ \mu-\frac{1}{2} m \omega_z^2 z^2 + V_0
e^{-(\frac{z}{\sqrt{2}l_z})^2} \}. \label{number}
\end {eqnarray}
By taking the integral in Eq.(\ref{number}), we find an analytical
expression for total number of atoms in terms of chemical
potential, interaction term, trap frequency and strength of the
dimple potential:
\begin {eqnarray}
 N &=& \frac{l_z}{3\sqrt{2}U_0(V_0+m\omega_z^2l_z^2)^{3/2}} \{ 2\sqrt{(\mu+V_0)}(3\mu V_0 \nonumber \\
  &+& m\omega_z^2l_z^2(2\mu-V_0))
 \nonumber + 3 V_0(V_0+m\omega_z^2l_z^2) \nonumber \\
 &\times& \sqrt{(V_0+m\omega_z^2l_z^2)\pi}
  Erf(\sqrt{\frac{\mu+V_0}{V_0+m\omega_z^2l_z^2}}) \}. \label{number1}
\end {eqnarray}
Here $erf(x)$ is the error function. The error function can be
expanded in terms of $x$ where
$x=\sqrt{\frac{\mu+V_0}{V_0+m\omega_z^2l_z^2}}$.
$erf(x)=\frac{1}{\sqrt{\pi}}[2x - \frac{2 x^3}{3} + \frac{x^5}{5}
- \frac{x^7}{21} + O[x]^9]$. We insert this expanded term into
Eq.(\ref{number1}) and solve this equation numerically in order to
get chemical potential. Doing this we find the value of the
chemical potential $\mu=1.0800 \times 10^{-12}$ eV for m $=23$ amu
($^{23}\mathrm{Na}$), $\omega= 200$ Hz and $V_0=100 \hbar
\omega_z=6.5821\, 10^-13$ eV for $N=1\times10^6$ .

%

\section{Ultraslow light under EIT scheme}

We consider an EIT model for a gas of  $N$ three-level atoms
interacting with two laser beams in $\Lambda$ configuration. The
upper level is coupled to the lower levels via a strong drive
field with frequency $\omega_c$ and a weak probe field of
frequency $\omega_p$. At resonance the absorption of the probe
field  can be neglected. Weak probe beam propagates along the
condensate axis in the $z$ direction. Propagation of the ultraslow
wave packet in one dimensional inhomogeneous atomic condensate can
be described by \cite{eit3,tarhan}
\begin{equation} \label{eq:pulse}
\frac{\partial E}{\partial z} + \alpha(z) E + \frac{1}{v_g(z)}
\frac{\partial E}{\partial t} + i \,b_{2}(z) \frac{\partial^2
E}{\partial t^2}  = 0,
\end{equation}
where $\alpha(z)$ is the pulse attenuation factor, $v_g(z)$ is the
group velocity, and $b_{2}(z)$ is the group velocity dispersion.
The third order dispersion is found to be much smaller and
neglected \cite{tarhan}. EIT susceptibility\cite{eit2} of
Bose-Einstein condensate of atomic density $\rho$ is expressed as
$\chi=\rho\chi_1$ with
\begin {equation}
\label{chieit} \chi_1 = \frac{|\mu|^2}{\epsilon_0 \hbar}
\frac{{\rm i}(-{\rm i} \Delta + \Gamma_2/2)}{[(\Gamma_2/2 -{\rm
i}\Delta)(\Gamma_3/2 -{\rm i} \Delta) + \Omega_c^2/4]},
\end {equation}
where $\Delta=\omega-\omega_0$ is the detuning of the probe field
frequency $\omega$ from the atomic resonance $\omega_0$. In Eq.
(\ref{chieit}), $\Omega_c$ is the Rabi frequency of the control
field and $\mu$ is the dipole matrix element for the probe
transition. $\Gamma_2$ and $\Gamma_3$ denote the dephasing rates
of the atomic levels. The significant position dependent group
velocity for the optical pulse propagation can be calculated from
the susceptibility using the relation\cite{eit3}
\begin{eqnarray}
\frac{1}{v_g} &=& \frac{1}{c} - \frac{\pi}{\lambda} \frac{\partial
\chi}{\partial \omega}|_{\omega_{0}} \label{eq:vg}.
\end{eqnarray}
Here we take $\lambda=589$nm, and $\omega_0=2\pi c/\lambda$. In
Eq.(\ref{eq:vg}) group velocity depends on atomic density. As
mentioned in the introduction, the atomic density can be increased
by using a dimple potential \cite{pinkse}. An optical ultraslow
pulse propagates through the ultracold medium without absorption
due to the small imaginary part of the EIT susceptibility at
resonance.

\section{Results and Discussions}
\begin{figure}
\includegraphics[height=.2\textheight]{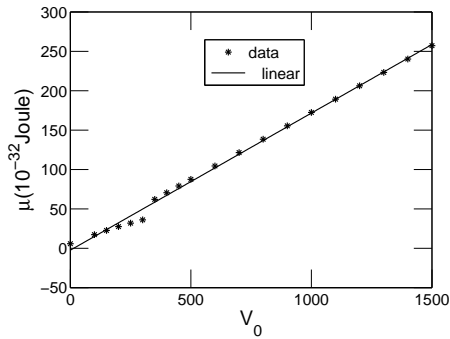}
\caption{The chemical potential $\mu$ vs strength of dimple
potential for $^{23}$Na the Bose-Einstein condensate of
$N=1\times10^6$ atoms. The dots show the chemical potential which
correspond to the strength of the dimple potential. Solid line
represents the linear fitting. Dimple potential ($V_0$) is scaled
by $\hbar\,\omega_z$. The parameters used are $M=23$ amu,
$a_{s}=2.75$ nm.} \label{fig1}
\end{figure}

We consider a gas of $N=1\times10^6$ $^{23}$Na atoms with
$\Gamma_3=0.5\gamma$, $\gamma=6 $ MHz, $\Gamma_2=6 \times 10^3$
Hz, and $\Omega_c=0.5\gamma$. We take accessible experimentally
parameters such as $\omega_{r}=350$ Hz and $\omega_{z}=100$ Hz, so
that peak density $\rho_0=1.56\times10^{20}$ 1/m$^3$. The
Eq.(\ref{number1}) is solved numerically in order to find the
chemical potential. We assume that all atoms are loaded into the
harmonic potential with a dimple therefore effective length of the
atomic medium becomes at the order $6 \mu$m. The density of the
condensate is mainly controlled by the dimple potential for
extremely deep dimple. We present the change of $\mu $ as a
function of $V_0$ in Fig.(\ref{fig1}). As the strength of the
dimple potential increases, chemical potential becomes larger.

\begin{figure}
\includegraphics[height=.2\textheight]{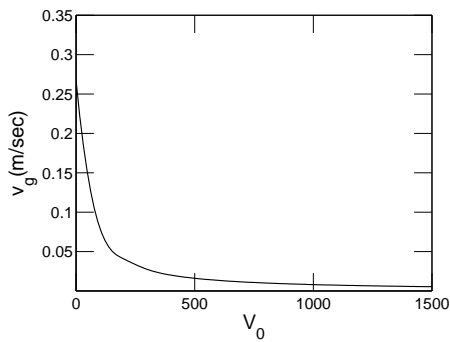}
\caption{Group velocity vs strength of dimple potential for
$\Delta=0$, propagating through a $^{23}$ Na Bose-Einstein
condensate under EIT scheme for $N=1\times10^6$ atoms. The solid
line shows the peak value of the group velocity vs strength of the
dimple potential. Dimple potential ($V_0$) is scaled by
$\hbar\,\omega_z$. The parameters are the same those of
Fig.{\ref{fig1}}.} \label{fig2}
\end{figure}

When light enters the condensate, its group speed exhibits a
dramatic slow down. Here we consider resonant probe pulse with
$\Delta=0$. Within the condensate region, at zero temperatures,
the group velocity remains approximately at the same ultraslow
value. Light rapidly accelerates to high speeds when it leaves the
condensate at the interface to thermal part. At extremely low
temperatures, $\rho_0$ saturates to Thomas-Fermi density. At zero
temperature, $vg_0$ can be calculated by $\rho_0$ which was
converted from Thomas-Fermi approximation. We can ignore the
spatial variations therefore density of the atomic system can be
taken as $\rho(z=0)=\rho_0$. In other words, at low temperatures
group velocity is determined by $\rho_0$ and group velocity
decreases rapidly with the increasing strength of the dimple for
small dimple strengths as seen in Fig.(\ref{fig2}). However, group
velocity decreases slowly at deep dimple. One can interpret the
decrease in the rate of slow down as follows: As $V_0$ becomes
larger, the dimple part predominates over the harmonic part of the
potential.

The group velocity is almost constant within the ultra cold atomic
medium. In this case we ignore the small contributions of modal
and waveguide dispersions and determine the group velocity, same
for both fractions, by assuming a constant peak density of the
condensate in the material dispersion relation.

So the delay time can be calculated approximately by the time
delay formula $t_{D}=L_z/v_{g}$. Here $t_{D}$ is the time delay of
the ultraslow pulse, $v_{g}$ is the group velocity in the $z$
direction and $L_z$ is the axial length of the condensate. Here
axial length of the condensate ($L_z$) can be taken as: $L_z=2R$
where $R$ is the Thomas-Fermi axial radius which is given by
$\sqrt{2( \mu+V_{0})/m\omega_z^2}$. Therefore we find large group
delays at the order $1$ msec in an atomic Bose-Einstein condensate
in a harmonic trap with a deep dimple potential in which the
strength of the deep dimple potential is $V_0=1500
\hbar\,\omega_z$.

\section{Conclusion}
We have explored the propagation of ultra slow light through a
Bose-Einstein condensate in a harmonic trap  with a dimple
potential. We have investigated the effect of the dimple potential
on the group velocity and the time delay. As the strength of the
dimple potential increases group velocity becomes smaller.
However, at a critical value of the dimple group velocity becomes
approximately constant. As a result, time delay can be increased
by means of an atomic Bose-Einstein condensate in a harmonic trap
with a dimple potential.

\acknowledgements

We thank Z. Dutton for valuable and useful discussions. H. U.
acknowledges support by  TUBITAK TBAG (108T003). D.T. was
supported by TUBITAK-Kariyer grant No. 109T686.

\end{document}